\documentclass[aps,prb,groupedaddress,amsmath,amssymb,showpacs,eqsecnum,
twocolumn, floatfix]{revtex4}
\usepackage{graphicx}

\begin{document}

%\preprint{}

\title{Superconducting proximity effects in  
metals with a repulsive pairing interaction} 


\author{Oriol T. Valls}
\email{otvalls@umn.edu}
\altaffiliation{Also at Minnesota Supercomputer Institute, University of Minnesota,
Minneapolis, Minnesota 55455}
\affiliation{School of Physics and Astronomy,
University of Minnesota, Minneapolis, Minnesota 55455, USA}
\author{Matthew Bryan}
\email{bryan175@umn.edu}
\affiliation{School of Physics and Astronomy,
University of Minnesota, Minneapolis, Minnesota 55455, USA}
\author{Igor \v{Z}uti\'{c}}
\email{zigor@buffalo.edu}
\affiliation{Department of Physics, University at Buffalo, Buffalo, New York
14260, USA}
%\thanks{}

\date{\today}

\begin{abstract}
%iz4 reword
%otv4 I like your changes,some minor rewording
Studies of the superconducting proximity effect in  normal 
conductor/superconductor $(N/S)$ junctions almost universally assume no 
effective electron-electron  coupling
in the $N$ region. While such an approximation leads to 
a simple description of the proximity effect, it is unclear how it could be 
rigorously justified. We reveal a much more complex picture of the proximity 
effect in  $N/S$ bilayers, where $S$ is a clean $s$-wave BCS superconductor
and $N$ is a simple metal with a repulsive effective electron coupling.
We elucidate the proximity effect  behavior using a 
highly accurate method to self-consistently solve the Bogoliubov-deGennes 
equations. We present our results for a wide range of values of the interface 
scattering, the Fermi wave vector mismatch, the temperature, and the ratio 
$g$ of 
the effective interaction strengths in the $N$ and $S$ region. We find 
that the repulsive interaction, represented by a negative $g$, strongly 
alters the signatures of the proximity effect as can be seen in the spatial 
dependence of the Cooper pair amplitude and the pair potential, as well as 
in the local density of states near the interface.

\end{abstract}

%\keywords{}

\maketitle

\section{Introduction}
\label{intro}
%otv3 rewritten again %iz4 ok
For nearly eight decades it has been recognized that  superconducting 
properties can leak out from a superconductor into a neighboring 
metallic region\cite{Holm1932:ZP,parks,Wolf:1985} which by itself
would not be superconducting.
This phenomenon is known as the superconducting 
proximity effect.
That superconductivity
can penetrate into a non-superconductor for a long distance,\cite{parks} 
has fascinated the condensed matter community ever since
this was discovered.

The main signatures of the proximity effect include the 
penetration of the Cooper pairs, with  associated phase coherence, into the 
non-superconducting region, and the  
suppression of the pair potential (the superconducting order parameter)
in the superconductor, near the interface.
Important insights in the proximity 
effect\cite{Pannetier2000:JLTP,Zagoskin:1998} 
are provided by its connection to the process of Andreev-Saint James 
reflection.\cite{Deutscher2005:RMP,pdgstj,Andreev1964:SPJETP,%
Griffin1971:PRB,btk,Bruder1990:PRB}
%It is instructive to recall a simple case of normal metal/superconductor 
%$(N/S)$ junction. 
%iz4 we need to define N and S, that existed in the previous version
An incident electron approaching %an 
a normal metal/superconductor
($N/S$) interface from 
the $N$ region can be reflected as a hole, resulting in the transfer of a 
Cooper pair into  the $S$ region. This is a phase-coherent scattering process 
in which the reflected particle carries information about both the phase 
of the incident particle and the macroscopic phase on the 
superconductor.\cite{Lambert1998:JPCM} 
Thus, Andreev reflection 
is responsible for introducing phase coherence
in the normal region. %\cite{Zagoskin:1998} 
Since this reflection is a two-particle process, 
it is plausible to conclude 
that the proximity effect will  be also weaker whenever
this anomalous reflection is 
suppressed, as e.g., 
in low-transparency $N/S$ junctions. 

Very impressive  advances in the fabrication of superconducting 
junctions (including atomically-flat interfaces\cite{Bozovic2002:N})
have  in recent years stimulated extensive 
experimental and theoretical studies of the proximity effect.  
%some basic questions are yet to be understood. 
For example, many  
recent efforts have focused on elucidating proximity effects in junctions 
including 
ferromagnets\cite{buzdin,Kontos2001:PRL,bergeret2,Demler1997:PRB,Radovic1991:PRB} 
or superconductors with unconventional (non $s$-wave) 
pairing symmetry.\cite{Deutscher2005:RMP,Bozovic2004:PRL,Asulin2004:PRL} 
However, significant challenges remain even for the studies of 
the proximity effect in a simple $N/S$ junction, where $N$ is the normal 
conductor and $S$ is a conventional BCS superconductor with 
phonon-mediated $s$-wave pairing symmetry. 
One such issue is that of the role
of the effective pairing interaction in the $N$ material.
In the simplest BCS version of the theory, the $S$ region is characterized by 
a  coupling constant $\lambda$ conventionally taken as positive for the
attractive case. In nearly all of the standard treatments of the
$S/N$ proximity effect this constant is assumed to vanish in the 
$N$ region.\cite{Ashida1989:PRB}  
This implies that the pair potential, %or, equivalently, the superconducting 
%order parameter,  
which enters in the underlying microscopic equations, 
would completely vanish in the $N$ region 
(although the pair amplitude would not)
for any choice of the $N$ region
and the $N/S$ interface. 
Yet, this zero coupling
assumption is hardly realistic: while the low-frequency
phonon mediated interaction is, on general grounds, always attractive,
the coupling $\lambda$ represents the difference between this attraction 
and the Coulomb pseudopotential, which is invariably repulsive. 
The balance of the two quantities may lead to a positive $\lambda$, 
leading to superconductivity, or a negative $\lambda$ but it is most 
unlikely that the two would exactly 
cancel. 
Indeed one would expect, 
in non-superconductors, negative values of $\lambda$ in roughly the same 
absolute value range of those found in superconductors.

%OTVR next paragraph rewritten to point indicated
This was already noted a long time ago\cite{pdg0} in a review article by
P.~G. de Gennes, and it 
was implicit in even earlier work.\cite{pdgstj,cooper} %OTVR1 "may" is silly
%OTVR1 cooper same vintage
%iz5 The idea was extensively followed up at the time:
%the comprehensive review article\cite{parks} mentioned
%above, discussed and reviewed many important
%aspects of proximity effect phenomena with emphasis on issues that could be 
%tackled at the time, given the
%experimental constraints implied by the quality of the samples 
%and interfaces then available, and taking into account also
%the limited  capacity of the existing %iz5  
%computers %iz5 then existing 
%which largely restricted
%theory to analytic methods.
%iz5 the above sentence to me looks too long, 
%Here is a possible rewording
The idea was extensively followed up at the time:
the comprehensive review article\cite{parks} mentioned
above discussed and reviewed many important aspects of proximity 
effect phenomena, with emphasis on issues that could be tackled at the time. 
The contemporary %OTVR1 did u mean that?
%iz6 I wanted to say about the constraints of that time (40 yrs ago)
%contemporary refers also to the present time; could it be confusing?
constraints were both experimental and theoretical, 
implied by the quality of the samples and interfaces
then available, the  absence of %OTVR1
suitable high-resolution probes such as the scanning tunneling %OTVR1 
microscope, and the limited  capacity of the existing computers, 
which largely restricted theory to analytic methods.
Among other topics, in Ref.~\onlinecite{parks} %OTVR2
%(previous too far) %iz6 OK
the effect on $T_c$ of an attractive or repulsive  interaction 
in the $N$ material 
was considered, 
%iz5 is mentioned to weak? should be put instead "considered"? %OTVR1 ok
with more
emphasis on the attractive case; a version of the so-called ``Cooper %OTVR1
limit''\cite{cooper} %OTVR1 
argument, for dirty superconductors and thin $N$ and $S$ 
layers, already given %OTVR1 too many discussed
in  Ref.~\onlinecite{pdg0}, %iz5 
was presented; %OTVR1  I think this is now straight
%to avoid discussed being repeated twice 
and a qualitative account of the energy gap
behavior at an $N/S$ interface was included.\cite{pages} 
As further
follow up to these %iz5 reviews additional work,
%does not sound smooth, maybe missing comma? 
reviews, additional work, 
including e.g. several %iz5attempts
%iz5 attempts sounds condescending
experiments\cite{ss1,ss2,ss3,lt13} that %otvR1 reword
used electromagnetic and transport properties to
estimate the pairing interaction and its sign, were subsequently %OTVR1
performed in the early seventies. %iz5 , 
Yet, despite
%the wide diffusion of %OTVR1 reword
%iz6 another rewording attempt %OTVR2 alternative accepted
the wide dissemination of
%iz6 if you do not like it consider replacing
%diffusion  with dissemination
%OTVR1 status of 
these reviews and the high reputation 
of their authors, 
%OTVR1 no the %iz5 needed? no
activity on this problem eventually
dwindled and relatively
little attention has been paid for many years to the
question of the influence of a {\it negative} value of $\lambda$ in the
$N$ material, on the proximity effect. 
The work of Refs.~\onlinecite{nagato} and \onlinecite{fauchere}, 
restricted to the quasiclassical  
limit, and that of Ref.~\onlinecite{pav} on critical currents
in the Cooper limit
are among the few exceptions. 
%OTVR end of rewrite 
The unspoken assumption elsewhere seems to have been 
that provided $\lambda$ is non-positive its value does not matter.
Yet, as pointed out already in Ref.~\onlinecite{pdg0}, such an assumption is
quite unreasonable. A repulsive interaction in the $N$ region will tend to
break the Cooper pairs coming from the $S$ region, which are responsible for the
proximity effect. The situation might be in some ways reminiscent of that 
found in the ferromagnet/superconductor %iz5 F now explicitly defined
$(F/S)$ proximity\cite{buzdin} effect for
a weak ferromagnet, where the range of the effect is reduced and 
signatures are found in the local density of states (LDOS) due to modified
Andreev reflection and Andreev-Saint James states.\cite{pdgstj} 
 
%OTVR rewrite 
In this work we reexamine %OTVR1 one this %iz5 long unsolved question 
this long-standing question %iz5 to tone it down, as Ref. 2 asks 
in a rigorous way.
We do this by solving the relevant microscopic equations, the Bogoliubov-de
Gennes\cite{bdg} equations, in a fully self-consistent way, using 
computational methods
recently developed and applied\cite{bvh,hvb} to study several
aspects of the $F/S$ proximity effects in clean systems with
smooth interfaces. 
%OTVR end of rewrite
We consider an $N/S$ 
bilayer in which each
layer is thicker than the superconducting coherence length, $\xi_0$, in the
$S$ material, and we study the properties of the system, focusing particularly
on both the Cooper pair amplitude and the pair potential 
as a function of position,
as well as on the LDOS near the interface. We study the
problem for several values of the interface scattering, the Fermi wave vector
mismatch between the two materials and, most important, of the value of the
{\it nonpositive} ratio $g$ between the effective pairing constant 
$\lambda_N$ on the $N$ side and the positive value $\lambda_S$
on the $S$ side. We find that the proximity effect markedly depends on $g$,
with the penetration 
of the pair amplitude
into $N$ being reduced as the absolute value of
$g$ increases. There is even a ``negative'' proximity effect: the
presence of a 
repulsive interaction in the
normal metal depletes the pair
amplitude in $S$ near the interface.
There are also signatures of the $g$ dependence on the LDOS as measured
very near the interface. Not surprisingly, these signatures are different 
from those apparent in the quasiclassical\cite{nagato} approximation, 
which cannot be accurate at very small 
length scales.\cite{quasi}
As a useful byproduct of our computations, we
find that it is erroneous, in the study of the 
$N/S$ proximity effect, to subsume the
separate effects  of interfacial scattering and wave vector mismatch into a
single effective parameter. This has been long known\cite{zv1,zv,stab} 
to be the case for $F/S$ interfaces but the situation  in the $N/S$ 
case was still unclear.

\section{Methods}
\label{methods}

%otv2 this paragraph totally rewritten
To study this problem, we 
solve self-consistently %iz4 rewording, equations appeared twice
the Bogoliubov-deGennes (BdG) equations,\cite{bdg} 
the relevant microscopic description for a clean system.
The geometry we consider 
consists of one normal metal slab of thickness $d_s$ %iz4 juxtaposed 
%to help some non-natives
%otv4  'next' could mean some distance away
juxtaposed to
a similar slab of
an ordinary BCS superconductor of thickness $d_S$. 
We assume a flat interface of  arbitrary transparency, 
infinite in the $x$- and $y$-directions, while the $z$-axis is normal
to the interface. 
The methods we use here have been presented and discussed 
elsewhere\cite{bvh,hvb,stab,kvb,us69,me}  and details need not be given
again here. 
%iz4 In the geometry just described 
In this geometry  
the BdG equations can be written as, %iz4 prefered by PRB, see your 69
%in the form:
%iz4 what about complex conjugate on Delta in (1,2) element ?
%otv4 it is real as seen in previous papers cited
%even if not needed, that could be said
\begin{equation}
\left( \begin{array}{cc}
H & \Delta(z) \\
\Delta(z) & -H \end{array} \right)
 \left( \begin{array}{c}
u_n^\uparrow(z) \\
v_n^\downarrow(z)
\end{array} \right)
= \epsilon_n
 \left( \begin{array}{c}
u_n^\uparrow(z) \\
v_n^\downarrow(z)
\end{array} \right),
\label{dgeq}
\end{equation}
%iz4 more traditional terminology, like in deGennes book
in terms of the spin-up quasi-electron, $u_n^\uparrow(z)$, and spin-down 
quasi-hole, $v_n^\downarrow(z)$,  amplitudes. 
%better to avoid repeated (( )) or use [( )]
%in terms of the spin-up quasi-particle ($u_n^\uparrow(z)$) and spin-down 
%quasi-hole ($v_n^\downarrow(z)$)  amplitudes. 
%iz4 reword, and reorder so that the used symbols are defined
%sooner after being used, similar to your PRB 69
%$\Delta(z)$, and it is to be determined self-consistently
%otv4 define things as they occur. 
Here $\Delta(z)$  is the pair potential 
(order parameter\cite{order}) which is to be determined self-consistently 
as explained below. The single-particle
Hamiltonian is
\begin{equation} 
H={k_z^2}/{2m}+\epsilon_\perp + U(z) - E_F(z),
\label{H}
\end{equation}
where ${k_z^2}/{2m}$ and $\epsilon_\perp={k_\perp^2}/{2m}$, 
denote the kinetic energy from motion in 
the $z$ and $x-y$ direction for parabolic bands,
respectively, $U(z)$ is a scalar potential and $E_F(z)$ 
represents the Fermi energies (band widths) $E_{FN}$, $E_{FS}$
%otv4
in the $N$ and $S$ regions.
We set $\hbar=k_B=1$, 
for Planck's and Boltzmann's constants. 
The variables $\epsilon_\perp$ are decoupled from the $z$ direction, %iz4 , added
but they affect the eigenvalues $\epsilon_n$. 
These variables are measured from the chemical potential. % as are
As pointed out 
already in many places,\cite{zv,bvh,hvb,stab,kvb} one
should not assume in this type of problem that the Fermi wave vectors
$k_{FN}$ and $k_{FS}$ (or, equivalently, $E_{FN}$, $E_{FS}$, 
as measured from the bottom of the bands) 
are the same in both materials. 
Thus, we introduce a 
dimensionless
mismatch parameter defined as 
\begin{equation}
\Lambda \equiv E_{FN}/E_{FS} = (k_{FN}/k_{FS})^2. 
\label{lambda}
\end{equation}
%otv3 reword below
We characterize the 
mismatch by this single parameter without the
additional introduction of
different effective masses in
the $N$ and $S$ regions,\cite{zd} which would not alter our findings.
%otv3 above last senetence could be left out? 
%iz4 let's leave it for now, it addresses possible questions why keep m's the same
%iz4 reword
In Eq.~(\ref{H}) we choose the scalar potential to describe the 
interfacial scattering
$U(z)=H_0\delta (z-z_0)$, where $z_0$ is the location of the
interface. The strength of this scattering is conveniently given in
terms of the dimensionless barrier strength $H_B \equiv  mH_0/k_{FS}$, 
which in the limit of no mismatch ($\Lambda=1$), 
coincides with the parameterization introduced in Ref.~\onlinecite{btk}. 
%The potential $U(z)$ describes interfacial scattering and we will take it
%of the form $U(z)=H_0\delta (z-z_0)$ where $z_0$ is the location of the

%otv4 stuf moved up
The BdG equations (\ref{dgeq})   
%The BdG equations 
above must be solved together with 
the self-consistency
condition:
\begin{equation}
\Delta(z) = \frac{\lambda(z)}{2}
{\sum_n}^\prime
\left[ u_n^\uparrow(z)   v_n^\downarrow(z)  +
       u_n^\downarrow(z)   v_n^\uparrow(z) \right]
\tanh\left(\frac{\epsilon_n}{2T}\right),
\label{self}
\end{equation}
%iz4 reword, define T
where $\lambda(z)=\lambda_N<0$ and
$\lambda(z)=\lambda_S>0$ in the $N$ and $S$  region, respectively,
the prime in the summation sign indicates that it is limited by the
usual Debye cutoff,  
and $T$ is the absolute temperature.

The procedures to diagonalize the BdG numerically while ensuring full
self-consistency\cite{selfc} have been explained  in the previous 
work mentioned above. 
Basically, one
chooses a convenient set of orthogonal functions, in this
and most cases it is appropriate %otv2
to take sine waves, and expands Eq.~(\ref{dgeq})
and (\ref{self}) in terms of that set. The required matrix elements
are the same as those used, for example, in Ref.~\onlinecite{me} in the
appropriate (nonmagnetic) limit. One assumes an initial form for the
function $\Delta(z)$ and then iterates the process until self-consistency 
is achieved, that is, the $\Delta(z)$ function obtained
from Eq.~(\ref{self}) is the same as the one input in Eq.~(\ref{dgeq}) at the
previous step.

\begin{figure}[tbh]
\includegraphics[width=3.0 in]{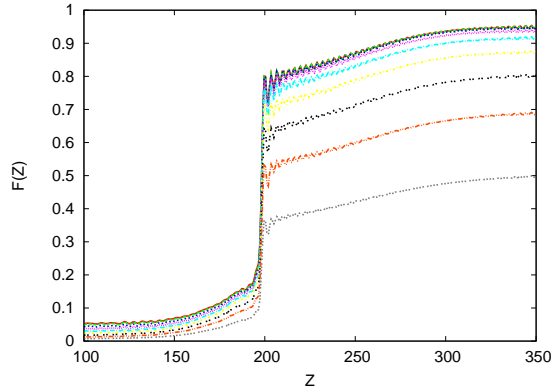}
\caption{(Color online) 
%iz2 fixing H_B,rewording Can we make it symmetric i.e. from Z 50 to 350?
%should we say how F(Z) is normalized?
The spatial dependence of the normalized (see text) %otv2
pair amplitude $F(Z)$. %otv2 across the N/S junction.
The dimensionless coordinate $Z$ is in units of the Fermi wave vector
in the $S$ region ($Z\equiv k_{FS} z$). The interface 
at $Z_0=200$ separates 
the $N$ region on the left with repulsive interaction ($g=-1/3$) from the 
$S$ region on the right which has attractive superconducting coupling.
The results are given for temperatures, expressed in terms of $T_c$, the
transition temperature of bulk $S$,
of   
$T/T_c$ of 0.01,  and 0.1 through 0.8, 
from top to bottom in the $S$ region. The interfacial scattering 
$H_B=1$ and the mismatch parameter
$\Lambda=0.5$ [Eq.~(\ref{lambda})] correspond to a low-transparency junction.
}
\label{fig1}
\end{figure}

Once self-consistency is achieved, one can directly examine quantities such
as the pair potential $\Delta(z)$ and the Cooper pair amplitude $F(z)$
(also known as the condensation amplitude) where, %iz4 PRB style : %otv3
\begin{equation}
F(z)= \Delta(z)/\lambda(z).
\label{f}
\end{equation} 
It is also very useful to examine, as a more
accessible experimental quantity,
the local density of states (LDOS) which is
obtained from tunneling experiments,
where the corresponding spectroscopic information can provided
by  scanning tunneling microscopy. 
We can express the LDOS $N(z,\epsilon)$ directly from the self-consistently 
calculated amplitudes ($u_{n\sigma}$, $v_{n\sigma}$) for the BdG equations 
(\ref{dgeq}) as,
\begin{equation}
\label{ldos}
N(z,\epsilon)=\sum_\sigma N_\sigma(z,\epsilon) 
= \sum_{\sigma,n} [u_{n\sigma}^2(z) 
\delta(\epsilon-\epsilon_n)+ v_{n\sigma}^2(z) \delta(\epsilon+\epsilon_n)], 
\quad \sigma = \uparrow, \downarrow. %kh2
\end{equation} 
One can integrate $N(z,\epsilon)$ %iz
over a wide region of $z$ to obtain the global DOS
or  over a very small region to obtain local results. Since our
methods are free of any quasiclassical assumptions, our results are
reliable even when the region examined is microscopic, i.e. of the 
order of the Fermi wavelength.

\section{Results}
\label{results}

The results of our calculations are described in detail in this section.
We will measure all the lengths in units of the Fermi wave vector $k_{FS}$
in the $S$ region, 
and define the relative dimensionless coordinate $Z\equiv k_{FS}z$.
The thicknesses of the $N$ and the $S$ regions are taken to be, in these
units, $D_N=D_S=200$ 
while the superconducting coherence length
(in the same units) is $\Xi_0=100$. Since $D_S=2 \Xi_0$, the pair potential
rises to very close to its bulk value at the far edge of the superconductor.
On the other hand, it does not decay all the way to zero in $N$: recall that
the %naive 
simplistic 
estimate\cite{parks} of the proximity depth is of order 
$T_F/T$ ($T_F$ is the Fermi temperature) 
in a clean system. It is obviously impractical for a numerical
calculation to make the $N$ slab thicker
than this value, and it is not necessary either for our study, which
focuses largely in the region near the interface. %The pair potential
%$\Delta(Z)$ and the pair amplitude $F(Z)$ are normalized to their bulk
%zero-temperature values in $S$. 
The values of $g=\equiv \lambda_N/\lambda_S$
studied are $g=0, -1/3, -2/3, -4/5$.  The values of the temperature are
given 
in terms of the ratio $T/T_c$ where $T_c$ is the bulk transition
temperature in $S$. Most of the data 
presented here are at relatively low temperatures ($T=0.1 T_c$) but for 
most of the 
values of $g$ we have obtained also results for reduced  temperatures 
of 0.01, 0.2, 0.3 and, in a few cases, up to 0.8 at 0.1 intervals. 
Values of the mismatch $\Lambda$  [Eq.~(\ref{lambda})] of 0.25, 0.5,
1, 2, and (in a few cases) 4 have been studied, while for the 
barrier parameter $H_B$ we have considered values of 0, 0.5, and 1. 

In Fig.~\ref{fig1} we show that the behavior of the pair amplitude 
$F(Z)$ as a function of temperature is as expected. 
In this figure we have used values of $H_B=1$ and
$\Lambda=0.5$ which correspond to  strong interfacial
scattering and high mismatch: hence a 
reduced proximity effect, 
as the $N$ and $S$ regions are weakly coupled. %nearly decoupled.
The value of $g$ is intermediate, $g=-1/3$.
$F(Z)$ is normalized so that its value in bulk $S$ material at $T=0$ would
be unity. 
Results for temperatures from nearly zero to 0.8 $T_c$ at approximately equal 
intervals are shown. The range of dimensionless distance from the left
edge $Z \equiv K_{FS}z$ includes one coherence length on the $N$ side
(at the left) and nearly all of the superconductor. 
The interface is (in all figures) at $Z\equiv Z_0=200$. We can 
see that $F(Z)$ rises towards the appropriate bulk value deep in $S$.
Because of the strong scattering and high mismatch, 
the profile
of $F(Z)$ near the interface is rather abrupt and the proximity effect
overall, quite week compared with other cases discussed below.
We see also that
the temperature dependence of the proximity effect is in this case not very
drastic [except as to the overall level of $F(Z)$] but appreciable.

\begin{figure}[tbh]
\includegraphics[width=3.0 in]{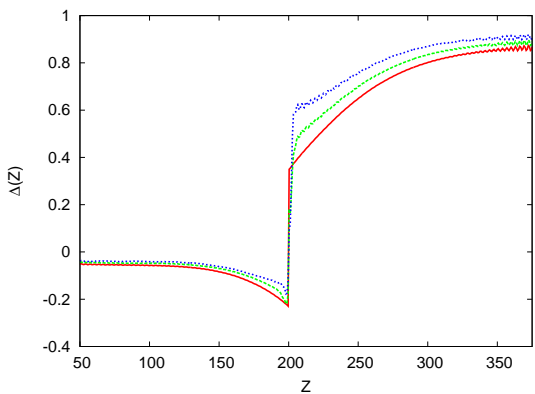}
\caption{(Color online) 
%iz2 rewording Can we make it symmetric i.e. from Z 50 to 350?
%should we say how Delta(Z) is normalized?
The spatial dependence of the pair potential $\Delta(Z)$ (normalized
in the same way as $F(Z)$) %otv2
for different
values of the interfacial scattering $H_B$. From top to bottom, the results
correspond to decreasing values of $H_B=1$ (blue), 0.5 (green), and 0 (red). 
$\Delta(Z)$ is calculated at low-temperature, $T=0.1 T_c$, in the 
absence of mismatch, $\Lambda=1$, and for a strong repulsive interaction 
in the $N$ region, $g=-2/3$.
}
\label{fig2}
\end{figure}

In the next figure, Fig.~\ref{fig2}, we discuss the influence of the interfacial
scattering parameter $H_B$ at a relatively high value of $|g|$ 
($g=-2/3$) and, for clarity,
in the absence of mismatch ($\Lambda$ =1). The quantity plotted this time is the
pair potential $\Delta(Z)$ (the ``order parameter'') as a function of $Z$.
$\Delta(Z)$ is normalized in the same way as $F(Z)$. 
Results for the three values of $H_B$ studied are shown.
%iz2 problems it is written as Delta in contradistinction w/ itself
%Of course, and in
%contradistinction with the pair potential, $\Delta(Z)$ is negative in the N
%material (except at  $g=0$). 
In contrast to $F(Z)$, the pair potential in the $N$ region will be negative
for a repulsive interaction ($g<0$).
The negativity of $\Delta(Z)$ in $N$ causes it to
abruptly jump at the interface. We see that this jump increases with %otv2 !!
$H_B$. This is as one would expect since higher barrier scattering
isolates the $S$ from the $N$ material and leads to increased pair potential
in $S$ near the interface and to less leakage of Cooper pairs on the $N$ side.

\begin{figure}[tbh]
\includegraphics[width=3.0 in]{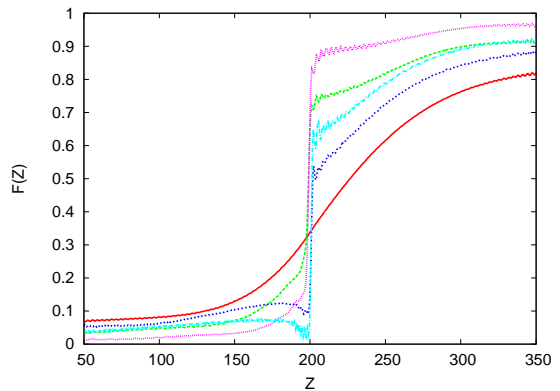}
\caption{(Color online)  
The spatial dependence of the pair amplitude $F(Z)$ for five different
mismatch values $\Lambda$ =1/4 (purple), 1/2 (green),
4 (cyan), 2 (blue), and 1 (red), from top to bottom on the right side.
The results are given for $H_B=0$, $g=-1/3$ at $T=0.3 T_c$.
}
\label{fig3}
\end{figure}

In a complementary way, 
for this two-parameter description $(H_B$, $\Lambda$) of  the $N/S$ interface,
we show in Fig.~\ref{fig3}, 
results for the effect of the mismatch
parameter $\Lambda$ at nonzero $g$ ($g=-1/3$ in this case) at $H_B$=0.
This time the quantity shown is $F(Z)$ (at smaller $|g|$ the $N$ side
the curves for $\Delta(Z)$ are harder to see). Results are shown for five
values of $\Lambda$ ranging from 1/4 to 4. 
It is evident that the proximity
effect is dramatically enhanced when the absence of interfacial scattering
is combined with the absence of mismatch ($\Lambda=1$), i.e., in the 
Cooper limit. %iz5
Indeed, in this
case (the (red) continuous curve) the behavior of $F(Z)$ at the interface is
very smooth, without any sign of of abruptness. In the scale shown in
the figure, and even at the finite temperature ($T=0.3T_c$) in the figure,
the pair potential  appears to settle into a constant on the $N$ side. This
is of course not true: it keeps decaying but very slowly since the proximity
depth is, in this case\cite{parks} much longer than the range shown in the
figure and, indeed, longer than the numerical sample size.

In the previous two figures, Figs.~\ref{fig2} and \ref{fig3}, we have considered
separately the influence of the parameters $H_B$ and $\Lambda$. 
This is actually necessary: it is often assumed that these two parameters
can be subsumed into a single  
parameter $Z_\mathrm{eff}$ that characterizes the effective barrier strength.
In our notation, $Z_\mathrm{eff}$\cite{bt} would be related 
to $H_B$ and $\Lambda$ as, %iz4 :
\begin{equation}
Z_\mathrm{eff}=\Bigl[\frac{H_B^2}{\Lambda^{1/2}}
+\frac{(1-\Lambda^{1/2})^2}{4\Lambda^{1/2}}\Bigr]^{1/2}.
\label{zeff}
\end{equation}
It is instructive to relate this $Z_\mathrm{eff}$ to the normal state junction
transparency, i.e., the transmission coefficient of an $N/N$ junction
at normal incidence\cite{btk,bt,zd}
\begin{equation}
T_{NN}=1/(1+Z_\mathrm{eff}^2),
\label{tnn}
\end{equation}
implying that $Z_\mathrm{eff}=0$ (or, equivalently, $H_B=0$ and $\Lambda=1$) correspond
to a completely transparent $N/N$ junction, while $Z_\mathrm{eff} >> 1$ would correspond
to a very low-transparency (tunneling) limit.
While such a single parameter ($Z_\mathrm{eff}$) interface description  offers a 
simplified approach and has been widely used, it may  lead to even  
{\em qualitatively} incorrect trends, as compared to the
correct description in which
the effects of $H_B$ and $\Lambda$ are considered separately. This has
been discussed in the context of $F/S$ junctions\cite{zv,bv} where it was  
%iz4 possibly 
possible
to find an example in which the mismatch can actually {\em enhance} the junction
transparency.\cite{zv1,zd,igor,zhu,zeng,feng} 

\begin{figure}[tbh]
\includegraphics[scale=0.6]{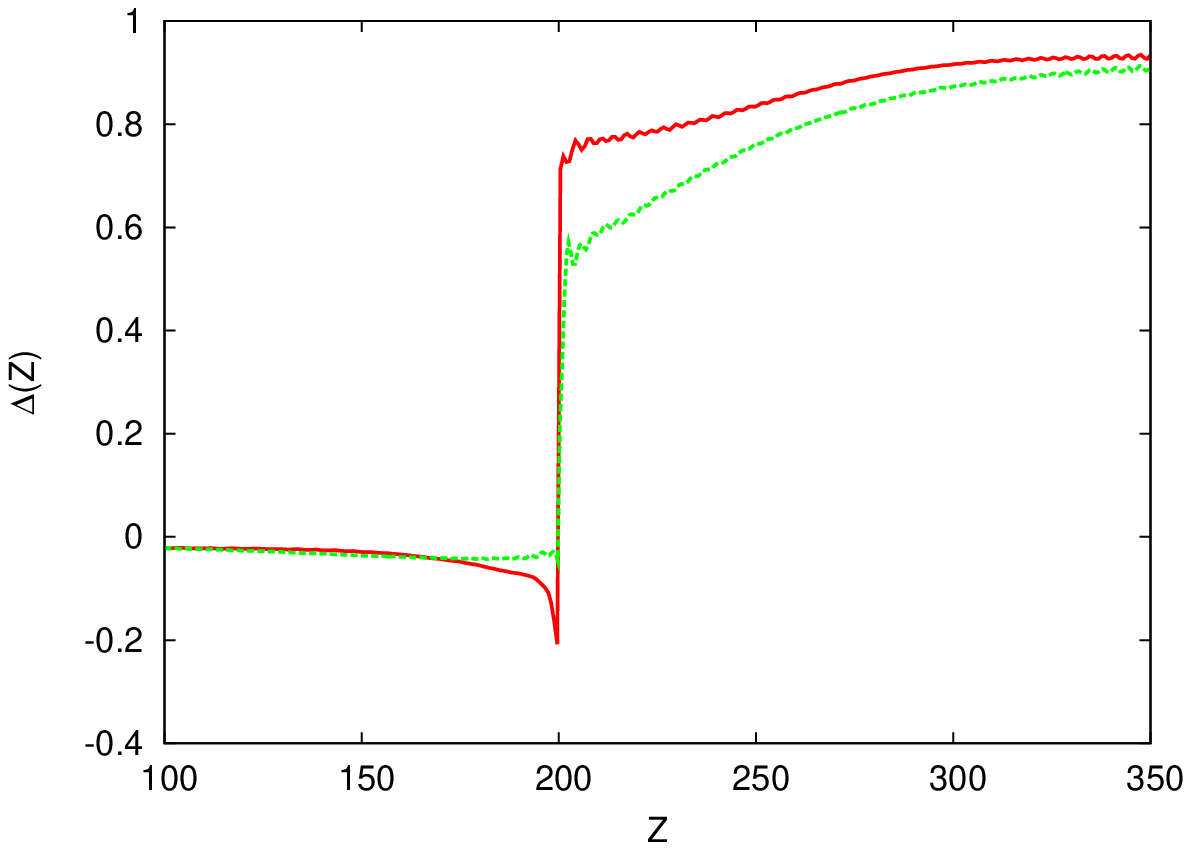}
\includegraphics[scale=0.6]{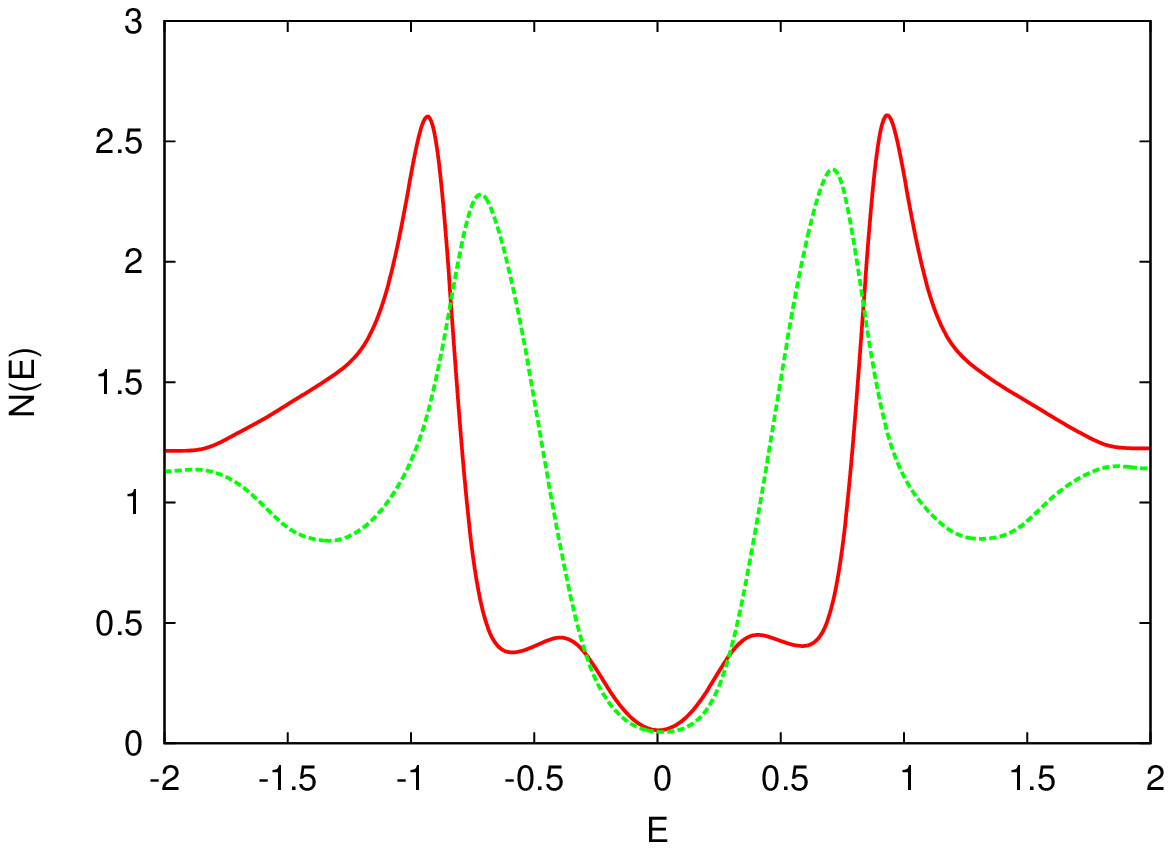}
\caption{(Color online)
Comparison of results for two different pairs of interfacial scattering and 
mismatch values, ($H_B,\Lambda$)=(0,1/2) and (0,2), which lead to the same 
effective barrier strength $Z_\mathrm{eff}$ 
[Eq.~(\ref{zeff})]. 
The %iz4 first 
%otv4 in 2 column version they'll be on top of each other
first panel shows 
$\Delta(Z)$. The top (red) curve at large $Z$ is for the first set of values, 
the other (green) curve is for the second set. The right panel shows the local 
density of states (LDOS) $N(E)$ near (see text) $Z=205$. 
The energy $E$ is in the units of the bulk 
zero-temperature superconducting gap $\Delta_0$. $N(E)$ is normalized  
to its value in the normal state of the bulk $S$ material (see text). %otv2 
The curve with the higher
(red) peaks is the for first set of values, the lower (green) 
peaks are for the second set. The results in both panels,
evaluated at $T=0.1 T_c$ and $g=-1/3$, 
clearly show that $Z_\mathrm{eff}$ alone can not describe the proximity 
effect or LDOS in $N/S$ junctions.
}
\label{fig4}
\end{figure}

Our results reveal that $Z_\mathrm{eff}$ alone is also inadequate to describe the 
proximity effects even in $N/S$ junctions. Equivalently, just characterizing
such junctions with either the corresponding transmission or reflection
coefficient, as frequently done\cite{nagato}, does not 
provide an unambiguous description.  
This can be clearly seen 
%for the pairs of %otv2 some rewording here
in Fig.~\ref{fig4} where two chosen pairs of $(H_B, \Lambda)$ values
[(0,1/2) and (0,2)] both
yielding the same $Z_\mathrm{eff}=0.174$ (an intermediate
value  also  equivalent  %otv2
to $H_B=0.174$ 
and no mismatch, $\Lambda=1$) produce
strikingly different results for both the spatial 
dependence of the pair amplitude and the local DOS. % (and thus the pair potential).
The two curves for $F(Z)$ display very different suppression near 
the interface in 
the $S$ region and different
decay in $N$. Analogous differences could also be observed for %(0,1/4) and
%otv2(0,4) results 
other pairs leading to the same or similar $Z_\mathrm{eff}$.
Clearly, the difference of calculated pair amplitudes which correspond to the
same $Z_\mathrm{eff}$ also implies that other quantities such as the pair potential 
and the LDOS must also be inequivalent for the same $Z_\mathrm{eff}$. 
This nonequivalence is clearly shown for the LDOS near the
interface in the second panel of Fig.~\ref{fig4}
for the same pairs of values (0,1/2) and (0,2). There we show the LDOS 
averaged over a region %otv2
five $Z$ units wide centered at $Z=205$, i.e., near the interface.
The energy is in units of the bulk gap in $S$, $\Delta_0$, and the LDOS
is normalized to the value that it would have in an equivalent region in 
bulk $S$ material: that is, the quantity plotted would be constant
and unity in the normal state of bulk $S$ material. These
normalizations will be used in all LDOS plots below. 
The inequivalence is obvious. The results have
very different  peak positions
and their heights are not at all the same. %otv2
We find that this is true
in general. Only in a very crude sense are mismatch increases (that is,
values of $\Lambda$ different from unity) equivalent to increases in 
$H_B$, in that both lead to diminished proximity effect. 
%iz4 for now I would leave this out, the main thing you already correctly state above
%the rest reads a little like 20/20 hindsight
%otv4 I have reworded it
With a little reflection, and perhaps some hindsight,
one can realize that, since the nature of the surface scattering
(normal or Andreev) originating from the mismatch is not the same as
that arising from the barrier, the failure of this naive approximation 
should not have been so unexpected. %iz3 reword later

\begin{figure}[tbh]
\includegraphics[width=3.0 in]{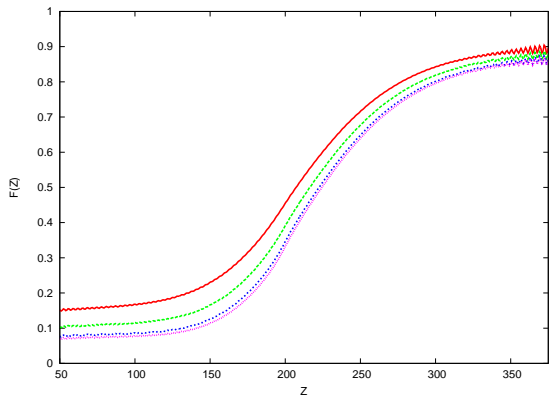}
\caption{(Color online) 
%iz2 rewording Can we show it symmetrically for Z from 50 to 350?
The spatial dependence of the pair amplitude $F(Z)$ for different values
of the repulsive interaction in the $N$ region: 
$g=0, -1/3, -2/3, -4/5$, from top to bottom.
The results are given for $T=0.1T_c$, $H_B=0$ and $\Lambda=1$. 
}
\label{fig5}
\end{figure}

Returning now to the basic question of interest, the effect of the
strength of the repulsive interaction in the $N$ region, %otv2 transition needed
we show in Fig.~\ref{fig5}  results for the pair amplitude $F(Z)$ 
for different values of the strength parameter $g$.
For the results in this figure the values of $H_B=0$ and $\Lambda=1$ correspond
to a transparent $N/N$ 
barrier [recall Eqs.~(\ref{zeff}) and (\ref{tnn})], 
i.e. a very strong proximity effect. Another signature of this high
transparency 
junction is the lack of Friedel oscillations, seen in the low-transparency
case in Fig.~\ref{fig1}.
$F(Z)$ does
not vanish even in the farthest region shown on the $N$ (left) side and indeed, as
expected, shows no sign of decay in the length scales shown. Towards the extreme
right of the figure, nearly two coherence lengths in $S$, $F(Z)$ 
approaches in all
cases its bulk value. However, looking in the regions near the interface
and in the $N$ region itself, we can clearly see how the proximity effect
is strongly affected by $g$: the pair amplitude decreases with increasing
$|g|$ not only in the entire $N$ region, but also in the $S$ region
within over one coherence length from the interface. Thus, as stated
in the Introduction, the influence of a negative $g$ pervades not only the $N$
material but also a thick region in the superconductor, where the Cooper pair
density is rather severely depleted.  

\begin{figure*}[thb]
%iz changing the order would help, 1st N, 2nd S-region and
%let's start with what was just shown in Fig.5 
%\includegraphics[scale=0.6]{fig5h0l1gL.eps}
%\includegraphics[scale=0.6]{fig5h0l1gR.eps}
%\includegraphics[scale=0.6]{fig5g23hL.eps}
%\includegraphics[scale=0.6]{fig5g23hR.eps}
\includegraphics[scale=0.6]{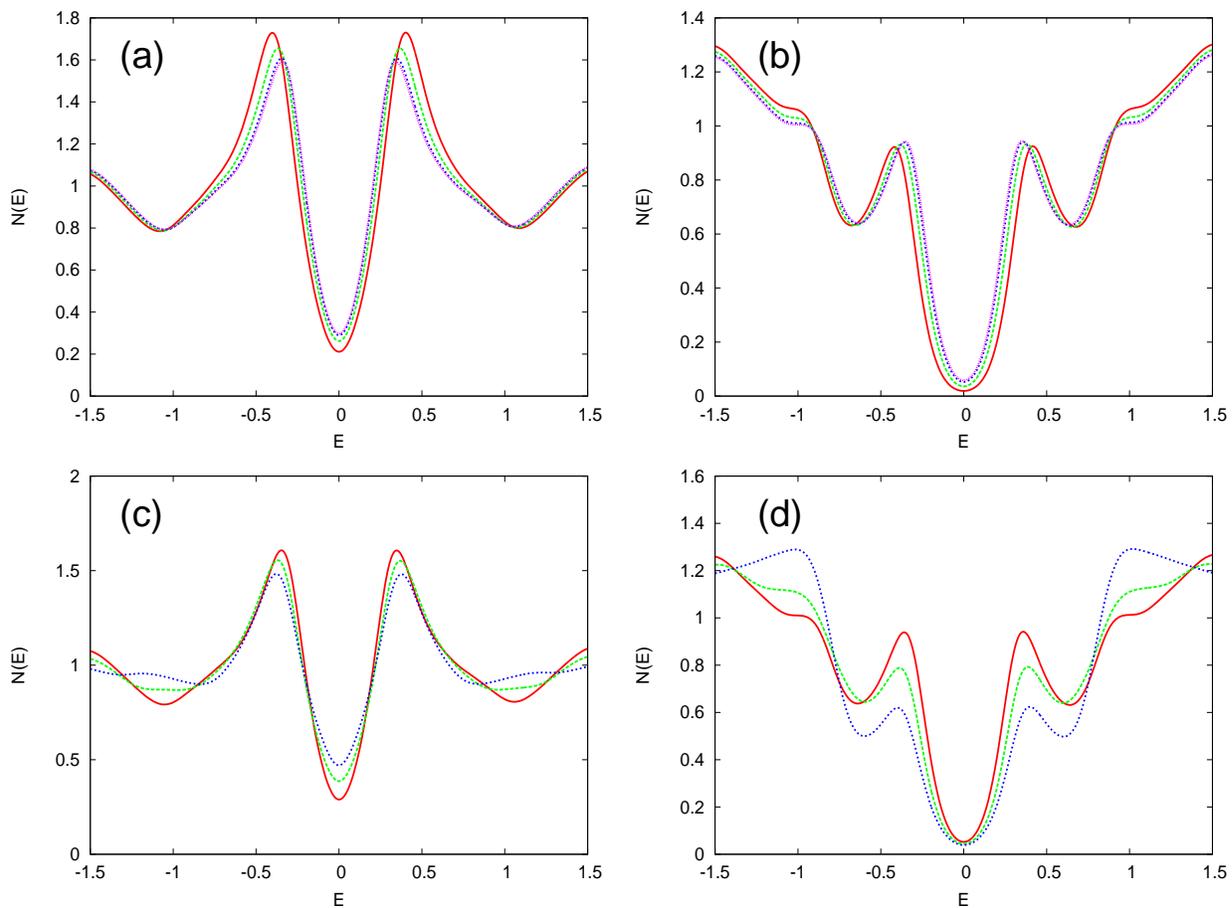}
\caption{(Color online) 
%it would help to put Z-value in each panel
%otv some small changes
The local density of states (LDOS) $N(E)$ [Eq.~(\ref{ldos})] in the
middle region of 
the $N$ and the $S$ regions (centered
at $Z$=100 and 300 respectively and averaged over
50 $Z$ units (see text) at $T=0.1 T_c$. %otv
%Panels (a) and (b) 
Panels (a) and (b) show $N(E)$ for the $N$ and $S$ sides, respectively, at
the same parameters used in Fig.~\ref{fig5}:  
$H_B=0$, $\Lambda=1$ and  $g=0, -1/3, -2/3 , -4/5$, corresponding 
to curves (red, green, blue and purple) for which $N(0)$ increases with $|g|$.
%Panels (c) and (d) 
Panels (c) and (d) similarly show the
LDOS for the parameter values used in Fig.~\ref{fig2}: 
$g=-2/3$, $\Lambda=1$, and three values of $H_B$, $H_B=0, 0.5, 1$ (red, green
and blue), LDOS peaks near $E=0.5$  are lower with increasing $H_B$.
}
\label{fig6}
\end{figure*}

%iz please correct if inserted Z value is wrong 
%otv some corrections made
We next examine the LDOS in the middle of the $N$ and $S$ region,
about one coherence length away from each side of the  $N/S$ interface.
The results are averaged over a region of width 100
centered in the middle of the $N$ and $S$ regions ($Z=100$ and $Z=300$,
respectively). Energy and LDOS are normalized as explained
in connection with Fig.~\ref{fig4}. Panels (a) and (b) in
Figure~\ref{fig6} %otv(a) and (b) 
show the LDOS evolution with $g$, for the same 
parameters as used in Fig.~\ref{fig5}. In both the $N$ 
(panel (a)) and $S$ (panel (b)) regions, $N(E)$ 
changes 
smoothly with $g$ without the 
appearance of any new features, as compared to the 
$g=0$ limit (in the absence of any repulsive interaction). In this high 
transparency limit ($H_B=0$, $\Lambda=1$), there are strong proximity effects 
as can be seen in  $F(Z)$ from Fig.~\ref{fig5}. We therefore expect and find
a $g$-dependent 
LDOS even one coherence length away from the interface,
in both the $N$ and $S$ regions. 
We note that the LDOS at $E=0$ is appreciably larger in the $N$ region. 
Increasing  $|g|$ leads to  larger  LDOS values near $E=0$, while the peak
near $E\pm 0.5$ moves to slightly larger values of $|E|$. At $E=1$
there is a vestige of a peak on the $S$ side, but not on the $N$ side. 
This is of course reasonable, since such a peak exists
in the bulk $S$, but not in the bulk $N$, material. %iz3 how? %otv3 needed?
In the panels (c) and (d) of Fig.~\ref{fig6} %(c) and (d) 
we consider the effect of the interfacial barrier 
strength on the  LDOS, averaged in the same way, for a fixed $g=-2/3$. 
With  increasing $H_B$
there is a suppression of the $N(E)$ peak at $E\approx \pm 0.5$ 
in both the $N$ and $S$ electrodes, 
but an increase in the peak at $E=1$ in the $S$ side, as the proximity
effect decreases.
Near $E=0$ the LDOS is enhanced only in the $N$ region. An increase
in $H_B$  diminishes the penetration of $\Delta(Z)$ in the $N$ region
and reduces  its depletion in  the $S$ region (see Fig.~\ref{fig2})
and this leads to the LDOS one coherence length away from the
interface looking rather 
%iz4 more bulk like as one increases $H_B$.
%otv4 I think you mean 'as' , 'as if' would not mae sense
more bulk-like as  $H_B$ increases.
This effect is more  marked if, in addition to increasing
$H_B$, one sets $\Lambda \neq 1$. %and
%thus lead to smaller energy variations in LDOS away from the $N/S$ interface. 
%Such trends can already be seen for given values of $H_B=0,0.5,1$. 
%iz Does this make sense, would be expect LDOS=const=1 away when Z_eff >> 1?
%otv it makes s ense, I have only changed the wording.

\begin{figure*}[tbh]
%\includegraphics[scale=0.6]{fig6h0l1F.eps}
%otv the Z range of this panel must be changed to be the same as the others. I
%otv will fix
%\includegraphics[scale=0.6]{fig6h0l1D.eps}
%\includegraphics[scale=0.6]{fig6h1l.5F.eps}
%\includegraphics[scale=0.6]{fig6h1l.5D.eps}
\includegraphics[scale=0.6]{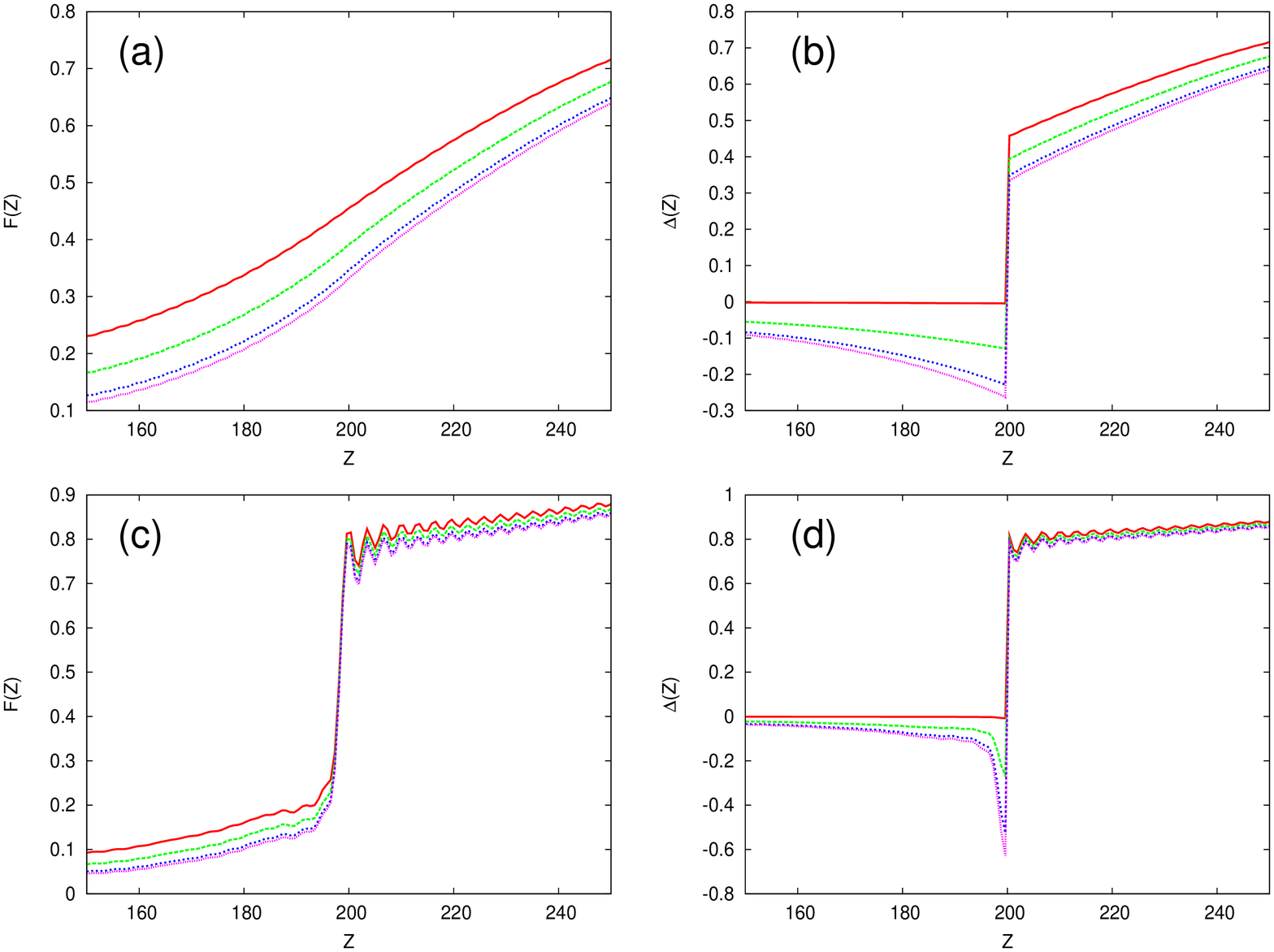}
\caption{ (Color online) The pair amplitude $F(Z)$ and the order parameter 
$\Delta(Z)$ plotted in regions near the interface for four values
of $g$, at $T=0.1 T_c$. Panels  (a) and (b)
correspond to $H_B=0, \Lambda=1$ (high transparency) and  
panels (c) and (d) 
to $H_B=1, \Lambda=0.5$ (low transparency). 
In all cases lower  values correspond to higher $|g|$.
(red, green, blue and purple curves correspond
to $g=0, -1/3, -2/3 , -4/5$, respectively)}
\label{fig7}
\end{figure*}
The effect of the repulsive interaction should be most pronounced
close to the $N/S$ interface. We focus 
next, therefore, on the spatial dependence of $F(Z)$ and
$\Delta(Z)$ near the
interface ($Z=Z_0=200$) as a function of $g$, 
in the low-temperature limit, $T=0.1 T_c$. 
Panels (a) and (b) in Fig.~\ref{fig7} %otv(a) and (b) 
correspond to the high-transparency limit
with $H_B=0$ and $\Lambda=1$ [recall that $Z_\mathrm{eff}=0$, from Eq.~(\ref{zeff})]. 
As seen already in Fig.~\ref{fig5} (the (a) panel
in Fig.~\ref{fig7} is a blow up of the interfacial
region of Fig.~\ref{fig5}), for every value of $g$ 
$F(Z)$ gradually increases with $Z$ and is smooth near the interface. 
In the limit of $g=0$ this behavior is well-studied.\cite{Ashida1989:PRB,me} 
On the other hand, right at the interface, there is a  
strong suppression of $\Delta(Z)$, 
as compared to the bulk value. In the $S$ region 
$\Delta(Z)$ and $F(Z)$ decrease very markedly with $|g|$ reflecting
that the repulsive interaction in $N$ induces a negative proximity
effect in $S$. 
In the $N$ region $\Delta(Z)$ 
is finite and negative
for $g<0$: this sign change is reminiscent of 
the pair potential behavior
due to the formation of $\pi$-states at interfaces with unconventional 
superconductors or F/S junctions.\cite{nagato,buzdin,Kontos2001:PRL,zv,us69,hu,%
%iz4 kashiwaya1,sengupta,wei,chen} 
kashiwaya1,sengupta,wei,Kashiwaya1999:PRB,Zhu1999:PRB,Kikuchi2001:PRB,chen} 
In this region, $|\Delta(Z)|$ increases with $|g|$ and decays away from the 
$N/S$ interface. For $g \leq -2/3$, the length scale for this decay exceeds %otv
%otv don't think so approximately a half of 
the superconducting coherence length.
In the other two panels, (c) and (d), of Fig.~\ref{fig7} %otv (c) and (d)
we consider, for comparison,
parameter values in the low-transparency limit
($H_B=1$, $\Lambda=1/2$). In contrast to the %otv Fig.~\ref{fig7}(a), 
other case, there
is a much sharper, nearly discontinuous rise of $F(Z)$ with 
a step-like behavior near the interface, for
every $g$. The rapid ($\sim k_F^{-1}$) oscillations 
that can be seen in the $S$ region for both $F(Z)$ and $\Delta(Z)$ are not 
numerical artifacts but represent Friedel-like oscillations 
induced by the sharpness of the interface,
which would not be seen in the quasiclassical treatment of this problem,
since such an approach would average over a $k_F^{-1}$ length scale. 
As compared to the high-transparency limit, the two main differences
that can be seen for $|\Delta(Z)|$ in Fig.~\ref{fig7} %otv (d):
are that it attains a much higher value in both the $N$ and $S$ regions next to the %otv3 
interface; and that it decays much faster in the $N$ region, 
away from the interface. %otv3 reworded below 
This decay of $|\Delta(Z)|$ is in general  nearly perfectly 
monotonic in both the  high- and low-transparency limits.
However, we have found some 
cases in which $F(Z)$ has a slight dip just inside the $N$ region,
next to the interface. This can occur 
when  $H_B$ is large (reflecting an
interface scattering potential, averaged over a  Fermi wavelength,
of order of the Fermi energy)  
or in the presence of large
mismatch (see e.g. Fig.~\ref{fig3}). In some cases 
this translates in $\Delta(Z)$
having a minimum slightly away from the interface. 
An example can be seen for $\Lambda=2$ 
in Fig.~\ref{fig4}. %otv3 [for $\Delta(Z)$].

\begin{figure}[tbh]
\includegraphics[width=3.0 in]{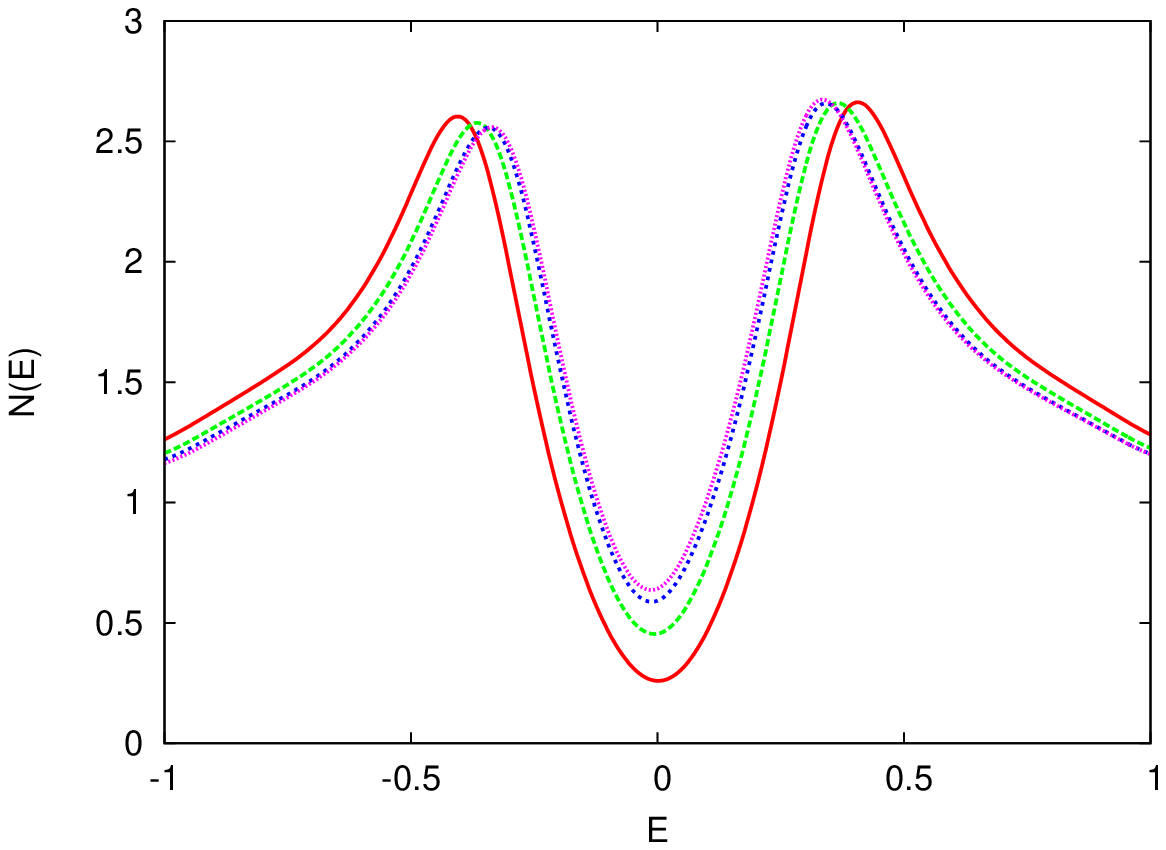}
\caption{(Color online)
%iz it would be better to show Z=199 instead
%otv it's nearly the same
The local density of states very near the interface (a region
one $k_{FS}^{-1}$ unit wide centered at $Z=201$, one unit from
the interface).
At  $Z=199$ the results are very similar. Results in this figure are for
$H_B=0$, $\Lambda=1$ and (from bottom to top near $E=0$) are for increasing
values of $|g|$ (red, green, blue and purple curves correspond
to $g=0, -1/3, -2/3 , -4/5$ respectively).
}
\label{fig8}
\end{figure}

%otv small changes below
We explore further the interplay of strong proximity effects and the repulsive
interaction in the $N$ region 
by considering the
interfacial LDOS at a region only one $Z$ unit wide centered at $Z=201$, %otv
just one hundredth of a coherence length from the $N/S$ interface,
in the $S$ region. %otv3
The  results shown in Fig.~\ref{fig8} are given for the 
high transparency $H_B$ and $\Lambda$ %otv
parameter values
used in Figs.~\ref{fig6} %otv(a), (b) 
and \ref{fig7}. %otv(a), (b).
Results at $Z=199$, just within the $N$ region, are very similar. %otv 
Comparing with the corresponding results in Fig.~\ref{fig6},
we see that the results near the interface are
qualitatively more similar %otv
to those well within (one coherence
length) the $N$ side than to those  well within the  $S$ side, 
and exhibit the same trends. Quantitatively, however, 
there are large differences: the peaks near $E=0.5$ for example, are
much more prominent at the interface. %otv new
This is not really surprising: the pair amplitude is already 
%iz4 drastically sounds already enough 
%very %otv3
drastically suppressed at the interface and (in this high transparency
case) it remains rather high in $N$ even one coherence length away.
The increased LDOS near the interface
arises at least in part from additional Andreev-Saint James states. %otv
Interfacial $\pi$-states have been shown to yield low-$E$ LDOS peaks as 
observed in $d$-wave superconductors and attributed to the formation of Andreev
bound states (ABS). 
Furthermore, even for $N/s$-wave-$S$ junctions with a repulsive 
interaction in the $N$ region, low-$E$ LDOS peaks have been predicted using 
analytical but not self-consistent
results (a step-function profile of $\Delta(Z)$) or employing
quasiclassical approximations to calculate $\Delta(Z)$.\cite{nagato,fauchere} 
%iz5 new 
For example, %OYVR1 reword
for a spatial profile of $\Delta(z)$ qualitatively %OTVR1
similar to the one
we calculated in Fig.~\ref{fig2}, the quasiclassical result of 
Ref.~\onlinecite{fauchere} (the inset of their Fig.~1) shows a very 
sharp zero energy  LDOS peak.
%iz5 rewording %OTVR more rewording
%In our results this zero energy peak is absent and we only
%see enhanced low-energy states. 
Our results show that, instead, the states at low energy
are enhanced but the zero-energy peak is absent. 
%rather than zero energy states. %otv a much stronger difference 
This is often found theoretically\cite{hvb,me} and experimentally\cite{court,sill}
in $F/S$ junctions.
These results could reflect 
that the quasiclassical approximation can alter the exact 
position of the low-$E$ interfacial LDOS peak for $g<0$.\cite{Walker1999:PRB}
%iz3  ref. I added is marginally relevant for this specific wording
We will further discuss the $Z$ dependence of the LDOS in connection
with Fig.~\ref{fig10} below. %otv

\begin{figure}[tbh]
\includegraphics[width=3in]{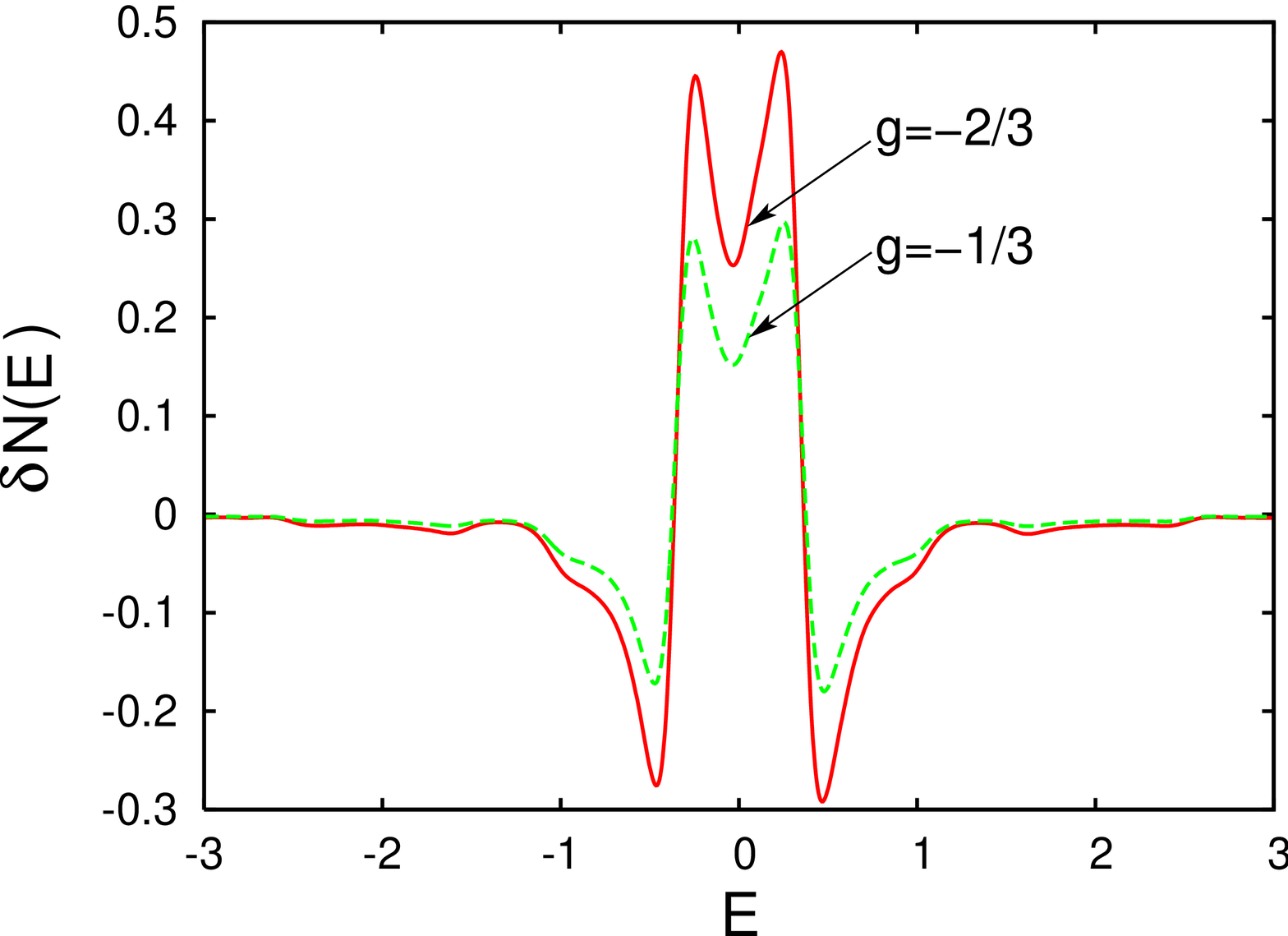}
\includegraphics[width=3in]{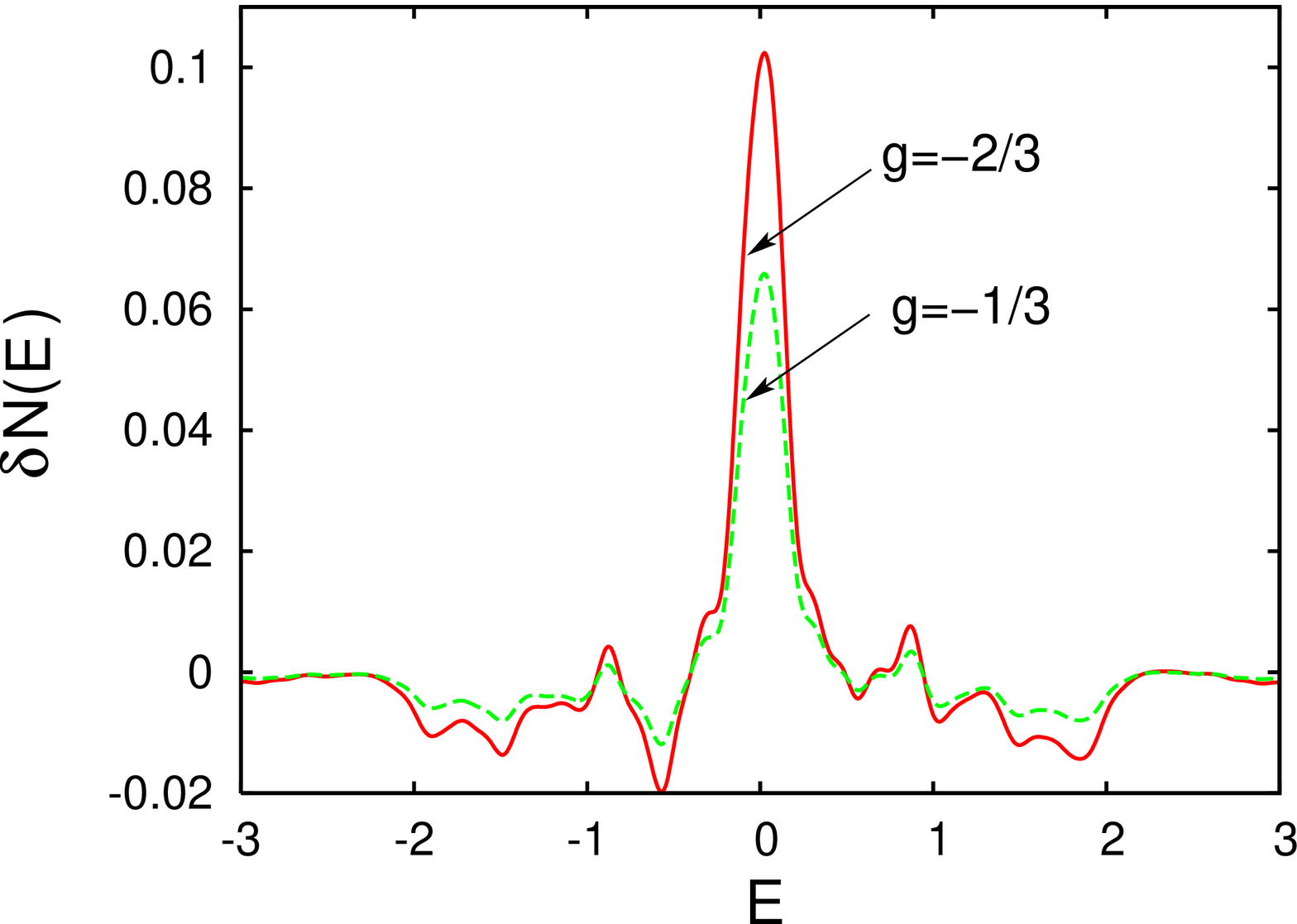}
\caption{Difference between the normalized LDOS, $\delta N(E)$,
near the interface ($Z=195$) 
at $g=0$ and  $g<0$. The first  and second panels correspond to a high-
and low-transparency interface with $H_B=0$, $\Lambda=1$, and 
$H_B=1$, $\Lambda=0.5$, respectively.
}
\label{fig9}
\end{figure}

To  study in more detail the influence
of $g$ on the LDOS
enhancement in the low energy region that we have observed,
we subtract the corresponding $g=0$  LDOS near the
interface from  its $g \neq 0$ value. These normalized
differences $\delta N(E)$  are
shown in  Fig.~\ref{fig9}, which corresponds to LDOS results averaged
over five $Z$ units centered at $Z=195$.
One can readily see that the effects of a nonzero $g$ are
quite significant in the low energy region.
The first panel corresponds to a high  
transparency junction, while the second
panel is for a low-transparency case. %otv not discussed before
We see that  the increase 
of the LDOS with $|g|$  is largest in the region
near $E=0$. The maxima are at $E=0$ in the low
transparency case and 
at nonzero $E$ at high transparency. 
%iz3 %otv3
The latter situation has been also found
to occur  in LDOS or conductance results and attributed
to  unconventional pairing with broken time reversal 
symmetry.\cite{Deutscher2005:RMP,zv,Covington1997:PRL,Fogelstrom1997:PRL,%
Zutic2005:PRL,Zutic1997:PRB}
These peaks 
for the LDOS difference are clearly %otv reworded
reminiscent of the %otv3 repeat LDOS 
peaks for the LDOS itself reported 
in quasiclassical studies for $g<0$
N/S junctions.%iz3 at $E=0$.  since they also show E>0 peaks
\cite{nagato,fauchere} 
In the quasiclassical approximation, the peak position was studied
as function of a reflection coefficient $R$ that assumedly 
characterizes overall the specific $N/S$ interface. The pitfalls
of using a single parameter for such purposes have been exposed
above, and the use of quasiclassical 
methods to study very narrow spatial regions at  the Fermi wavelength
level is obviously suspect. %otv 
It was expected that increasing $R$ would shift the LDOS
peak position from $E=0$ 
to finite $E$-values.\cite{nagato} 
Comparison with our rigorous, two parameter, results  
from %the right 
%otv panel of 
Fig.~\ref{fig9}, shows the 
opposite  trend for the differential LDOS peak, which 
moves to $E=0$ with decreasing transparency. %{\em increasing} $R$. 
%Not knowing that we show a differential 
%LDOS peak (rather than the LDOS peak itself), it would make 
Our low-transparency  differential LDOS 
results resemble the
frequently observed LDOS zero bias conductance peak, a signature of 
ABS in unconventional superconductors, which typically becomes narrower with 
the increase in $Z_\mathrm{eff}$ i.e., in $R$.\cite{zv}  
The trend 
we find, that a low-$E$ the interfacial $\delta N(E)$ increases with $g$, is
rather robust (we can see it for both high- and low-transparency and 
for a wide
range of temperatures, not just $T\leq 0.1 T_c$) and it is possible
that it could be used to directly identify $g<0$ experimentally. Such an
identification would be complicated by using the suitable $g=0$ LDOS
background subtraction. %iz this may even go after Fig. 10 %otv or conclusions

It is helpful to %otv3 consider alternative possibilities in
consider in
more detail the behavior of the LDOS as one
approaches the interface. While the existence of in-gap states 
is an inescapable consequence of the Andreev-Saint James
surface states, neither experiment\cite{court}
nor theory\cite{bvh,hvb} require that they be located
at zero $E$. %otv3
We  now examine here 
how the position of the low-$E$ LDOS peak  %otv2 could appear at a different 
depends on  the LDOS location in the $N/S$ structure
and how this dependence changes as $g$ varies,
%otv3  I have canceled the following as I did not think it was clear. Could be
%put back
%otv4 I put back some of it
and how this might correlate with %the spatial profile of $\Delta(Z)$,
the minimum in $\Delta(Z)$ being 
pushed away from the interface, as mentioned above. 
To elucidate this question, we consider the spatial evolution of the LDOS for 
both the $g=0$ and $g<0$ cases, as shown in Fig.~\ref{fig10}. Results are given 
for the LDOS, evaluated at regions centered  
at %otv2 additions here
distances 1, 3, 5, and 50 (in the usual units of $k_F^{-1}$), 
from each side of the $N/S$ interface, and averaged over  %otv2
a region of the same total width as the 
corresponding distance from the interface. As always, we normalize the
LDOS to the value it would have in an equivalent region of the bulk
$S$ material in its normal state. %otv2 I think it bears repeating.
%various distances from both sides of the $N/S$ interface. 
\begin{figure*}[tbh]
\includegraphics[scale=0.6]{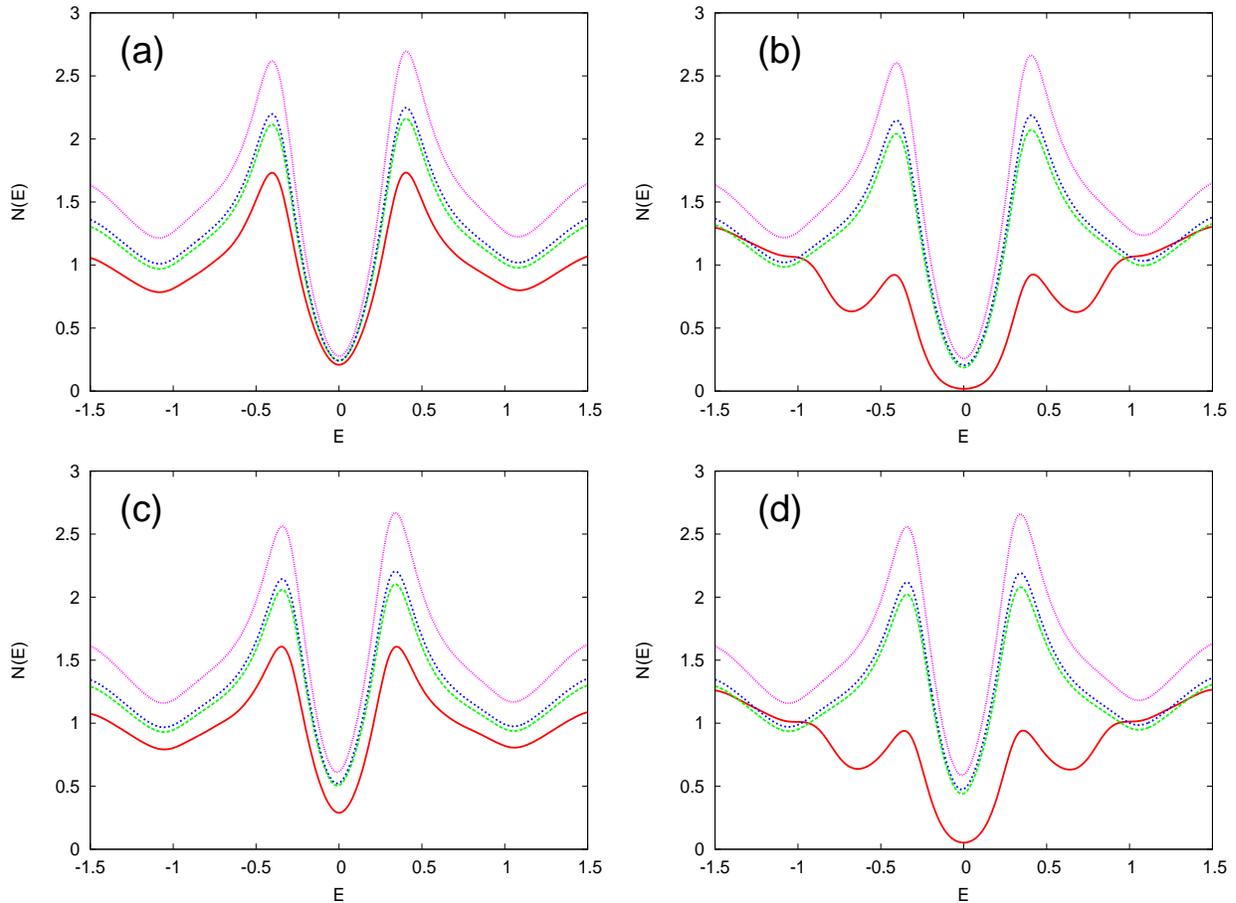}
\caption{ (Color online) The local density of states 
%iz2 some rewording and reordering %otv2 more of this
for various distances from the interface in the high transparency
limit ($H_B=0$ and $\Lambda=1$) at $T=0.1T_c$, 
Panels (a) and (b) show  results for the $N$ and $S$ regions of the
sample respectively, at $g=0$, while (c) and (d)
show results for $g=-2/3$.  
In all cases the ordering of curves, from top to bottom at $E=0.5$
corresponds to $Z$-regions at
distances, in  units of $k_F^{-1}$, 1, 3, 5, and 50 (purple,blue,
green and red curves), 
from each side of the $N/S$ interface, averaged over  %otv2
a region of the same total width as the distance from the interface.
%bottom to top at $E=1.5$, to a distance from the
%interface of 50, 5, 3, and 1 (in units of inverse $k_F$) as explained
%in the text.
%We have in all cases $T=0.1T_c$, $H_B=0$ and $\Lambda=1$. 
}
\label{fig10}
\end{figure*}
In the absence of a repulsive interaction ($g=0$, panels
(a), (b) for the $N$ and $S$ sides respectively), moving away 
from the interface reduces the the height of the 
low-$E$ LDOS peaks (below %otv3 
$E \approx \pm 0.5$) in both
the $N$ and $S$ regions. Near $E=0$ there is virtually no change in the 
$N$ region (consistent with $\Delta(Z)=0$, for $Z<200$) while, on the
other hand, there
is a marked decrease in the zero-$E$ LDOS as one moves
away from the interface in the $S$ region. %otv2

%otv3 split paragraph
Comparison of these findings with those
at $g=-2/3$ (panels (c) for the $N$ side and
(d) for the $S$ side), reveals a similar LDOS structure but
a different situation. %otv2
While there are no new zero-$E$ peaks, 
%otv2 that would be absent for $g=$. However, 
there are some clear differences. %otv3 which can be seen %otv2 reword below
Near $E=0$ in the $N$ region, moving closer to the interface does not
lead to new LDOS peak formation, but we can notice a clear enhancement in 
the zero-$E$ LDOS as the interface is approached
from either side, while %otv3 later, also on both sides
%of the interface, 
the height of the low-$E$ peaks increases very markedly as
the interface is approached. %otv3 new
On the $S$ side of the interface, the low $E$ peaks also increase
as the interface is approached, and move slightly 
%iz4 I think slightly alone is enough
%otherwise, if is is not visible, one would say almost do not move
%but perceptively
towards lower energies.
We have observed similar trends in the LDOS for other junction parameters at 
other distances from the interface. In all those cases studied, 
we %otv2 were not able to identify 
do not find a zero-$E$ peak as was associated with ABS 
in the quasiclassical studies.\cite{nagato} 
However, as can be seen from comparison of the $N$ regions for $g=0$ and $g=-2/3$, 
the repulsive interaction leads to an enhanced low-$E$ spectral weight. 
Subtraction of the LDOS, calculated close to the interface, for $g=0$ from 
that of $g=-2/3$ ((c) and (d) panels in Fig.~\ref{fig10})
would again lead to a peak in the differential LDOS near $E=0$, 
as shown in Fig.~\ref{fig9}. 
%iz3 reword
Thus, the %difference between
%these and the quasiclassical 
%results can be attributed
%to the superior spatial resolution of our methods,
%while the 
observable 
effect of $g$ %otv2
is confirmed. %otv2 should this be left to 'conclusions'?

\section{Conclusions} %otv3 edited throughout %iz4 OK
\label{conclusions} 
While the possible role of a repulsive effective electron-electron interaction
%in the normal material of ametal/superconductor $(N/S)$ junction,
%iz4 it may be better to say normal metal or normal conductor
%or it sounds that the superconductor is an abnormal material
%otv4 is this ok?
in the normal metal of a conductor/superconductor $(N/S)$ junction,
had already been noted in the early 
seminal work on the superconducting proximity effect,\cite{pdg0} 
subsequent studies  almost universally considered such interaction 
to vanish identically.
Perhaps the reason was that for the $N/S$ proximity effect
such a neglect   
leads to considerable simplifications. 
In the $N$ region, the pair potential $\Delta({\bf r})$ (the 
superconducting order parameter) vanishes identically and only 
the pair amplitude $F({\bf r})$ needs to be considered. %In principle, 
%then the decay of superconducting correlations away from the $N/S$ 
%interface, given by $F({\bf r})$, would be universal. %otv3 not true!
This leakage
of Cooper pairs in the $N$ region could then be 
approximately inferred by simply 
considering  Andreev reflection and a step-function pair potential,\cite{btk}
although this would involve also neglecting the depletion of
the pair potential in $S$.  
However, such assumptions, which would lead to the proximity effect
%otv3 this was wrong: H Lambda and xi_0 would still be important.
%would lead to a kind of universality of the proximity effect, that is
%completely independent of the $N/S$ interface properties 
being for many purposes independent on the choice of
$N$ material, could hardly be justified, theoretically or experimentally. 

In this work, we have carefully
and rigorously examined the various implications that 
the influence of a repulsive effective electron interaction in the   
$N$ region has
of the proximity effect
in an $N/S$ bilayer. In addition to the spatial variation of the pair 
amplitude, one also has
to study the decay of the finite pair potential in $N$, away from the $N/S$ 
interface, and its depletion in $S$.
Each of those spatial dependences are strongly affected both by
the effective interaction in $N$ and by the $N/S$ interfacial
properties. %otv3 below
In the $N$ material, they have opposite trends in the magnitude: 
$F(z)$ is suppressed
while $|\Delta(z)|$ is enhanced by a stronger repulsive interaction. 
In the superconductor, however, they are both depleted in the
same way.
This suppression of the pair potential near the interface is another signature
of the proximity effect. It directly depends on the strength of this repulsive 
interaction. %In some situations we even find that $|\Delta|$ does not decay
%otv3 weakmonotonically away from the interface. 
We also consider the dependence of the proximity effect
on interface scattering. Many studies of superconducting
junctions employ a single-parameter description (for example, 
using the corresponding normal state reflection or transmission coefficient)
for the interface properties. We show explicitly that this is clearly 
insufficient, even for the commonly used 
$\delta$-function model\cite{btk} of the interfacial barrier. 
More specifically, the nature of the proximity effect changes 
independently with {\em both} the strength of the interfacial barrier $H_B$ 
and the Fermi wave vector mismatch $\Lambda$ between the $N$ and $S$ regions.   

%In contrast to  previous quasiclassical studies of the influence of 
%repulsive interaction on the proximity effects, 
We have not found any
support for the formation of a zero bias (or near zero bias) peak in the
local density of states (LDOS) near the $N/S$ interface, usually attributed 
to the formation of Andreev bound states. We do find %iz4 of course, 
a plethora of in-gap states attributable to these bound states in
agreement with previous work. We find 
also similar zero $E$ features
for the differential LDOS, after performing a subtraction of the LDOS for
vanishing repulsive interaction. Such differential LDOS peaks become more 
pronounced with increasing  repulsive interaction and resemble the
zero bias conductance peaks studied extensively in unconventional 
superconductors.

%OTVR paragraph below rewritten
A challenge for  future work would be to identify specific materials and
systems where the explored proximity effects could be readily observed.
Our results show that, as might have been  expected,
the effects of a repulsive interaction are quantitatively important but not
qualitatively obvious: there is no 
simple and evident ``smoking gun''. The peaks we find
at small energies are in
$\delta N(E)$ not in the local DOS itself. However, careful quantitative studies
of the LDOS near the interface as a function of thickness should
be (see e.g. Figs.~\ref{fig8} through \ref{fig10}) revealing. 
Such studies have been technically possible for several years
and have been found useful (see e.g. Refs.~\onlinecite{court,sill})
in the study of $F/S$ systems. Furthermore,   
a repulsive interaction could be considered as a simple model for
a strongly-correlated $N$ region. Materials such as VO or Pd could be used
as the material forming the $N$ layer, 
as well as other materials that may have an enhanced susceptibility, close
to the Stoner instability. Such materials should
have a repulsive effective interaction.
%iz5 some additions %OTVR1 slight reword below
Another direction would be to further examine 
semiconductors as the $N$ region. Two classes of materials could be 
suitable candidates: ferromagnetic (III,Mn)V semiconductors, which %OTVR1
have revealed unusual Andreev  reflection in  $N/S$ %OTVR1
junctions,\cite{zd,Braden2003:PRL,Panguluri2005:PRB} and 
nonmagnetic narrow-band gap semiconductors. In the second class, 
it might be useful to focus on InAs-based semiconductors.
These materials offer high mobility and a suppressed Schottky barrier 
with $S$ region and  are already known for intriguing properties 
of Andreev reflection and proximity effects\cite{Schapers:2001}. 
Gating of such a two-dimensional semiconductor would offer a natural path 
to alter the strength of the repulsive interactions.  

%OTVR1 slight reword next few lines. %iz6 OK
Finally, recent experimental and theoretical advances could be used 
in the future to 
extend previous ideas about employing screening effects\cite{ss1,ss2,lt13} to 
extract the strength of repulsive interactions in the $N$ region. %OTVR1
With an applied magnetic field the proximity induced superconductivity 
in the $N$ region implies that there will also be a supercurrent and %OTVR1
a Meissner effect.  In the linear approximation for the Meissner 
effect (with the supercurrent proportional to the superfluid velocity),
the presence of $\pi$-states was associated with  paramagnetic  %OTVR1
instability at $N/S$ interface.\cite{fauchere} 
However, it has been shown that the region with suppressed %OTVR1
pair potential can also lead to an important {\it nonlinear} 
%OTVR2 repeated contribution to the Meissner %iz6 effect.\cite{Xu1995:PRB,Zutic1997:PRB} %OTVR1
contribution to the Meissner effect.\cite{Zutic1997:PRB,Xu1995:PRB} %Ref reorder
Revisiting the interplay of the screening response on the proximity 
effects could provide additional insights for probing repulsive 
interactions. 
%OTVR1 I think we should should we drop the next five lines 
%In particular, the effects of a time-dependent applied
%magnetic field lead to the generation of higher 
%harmonics~\cite{Zutic1998:PRB,Dahm1999:PRB} which can  
%be effectively combined with high-sensitivity penetration depth
%measurements\cite{Bidinosti2000:RSI,Prozorov2006:SST}.
Thus, there is reason to expect that additional ways
of obtaining and interpreting experimental data in order to extract the
effective pairing interaction will soon be available. %OTVR1 last coda  

\begin{acknowledgments}
%otv4
We thank  Paul H.~Barsic for many conversations regarding
his work and especially for  developing most of
the computer codes used to produce these result.
This work was supported in part by NSF-ECCS CAREER, ONR, and AFOSR.
I. Z. wishes to thank V. Lukic and J. Wei for discussions. 
\end{acknowledgments}


\begin{thebibliography}{99}
\bibitem{pages} In the context of the preceding sentences, pages 1012
to 1019 of Ref.~\onlinecite{parks} are particularly relevant.
\bibitem{order}
In some of the literature, 
e.g.  
Ref.~\onlinecite{pdg0}, the Cooper pair amplitude
(condensation amplitude) is loosely called the order parameter. 
We are careful here to maintain the distinction between
order parameter and pair amplitude according
to  Eqs.~(\ref{self})
and (\ref{f}) below.
\end{thebibliography}

\end{document}